\documentclass[twoside]{article}

\def\be{\begin{equation}}
\def\ee{\end{equation}}

\def\BibTeX{{\rm B\kern-.05em{\sc i\kern-.025em b}\kern-.08em
            T\kern-.1667em\lower.7ex\hbox{E}\kern-.125emX}}

\usepackage{graphicx}

\begin{document}
\sloppy
\twocolumn[{
{\large\bf NON-EQUILIBRIUM PHASE TRANSITIONS IN ULTRARELATIVISTIC\\ NUCLEAR COLLISIONS}\\

{\small B.~Tom\'a\v{s}ik$^{a,b}$ (boris.tomasik@umb.sk), G.~Torrieri$^c$, I.~Melo$^d$, P.~Barto\v{s}$^a$, 
M.~Gintner$^{a,d}$, S.~Kor\'ony$^a$, I.~Mishustin$^e$\\
$^a$ Univerzita Mateja Bela, Bansk\'a Bystrica, Slovakia, $^b$ \v{C}esk\'e vysok\'e u\v{c}en\'{\i} technick\'e,
Praha, Czech Republic, $^c$ Universit\"at Frankfurt, Frankfurt/Main, Germany, $^d$ \v{Z}ilinsk\'a Univerzita,
\v{Z}ilina, Slovakia, $^e$ Frankfurt Institute for Advanced Studies, Frankfurt/Main, Germany}\\


{\bf ABSTRACT.} Highly excited nuclear matter created in ultrarelativistic
heavy-ion collisions possibly reaches the phase of quark deconfinement. It
quickly cools down and hadronises. We explain that the process of hadronisation 
may likely be connected with disintegration into fragments. Observable 
signals of such a scenario are proposed.  \\
}]

Ultrarelativistic nuclear collisions are probed with the aim to create and study nuclear 
matter under most extreme conditions ever created in laboratory \cite{boris-ksf}.
In collisions at the Relativistic Heavy Ion Collider (RHIC) of Brookhaven National Laboratory
and in future collisions at the LHC (CERN) matter in state with deconfined quarks and 
restored chiral symmetry is 
produced. Due to initial conditions with longitudinally fast moving nuclei and strong inner 
pressure the created deconfined bulk matter expands very quickly. It cools down, returns 
into hadronic phase, and disintegrates into individual final state hadrons in a short time 
period of the order 10~fm/$c$, or $10^{-22}$~s. It is less clear whether deconfined and chirally
restored matter is 
produced in collisions at lower energies, like those at CERN's 
Super-Proton Synchrotron (SPS) (see e.g.\ discussion in \cite{boris-ksf}). In any case, at 
lower collision energies matter is produced with higher baryochemical potential. 

\begin{figure} [h,t]                  
\begin{center}                        
\includegraphics[width=80mm]{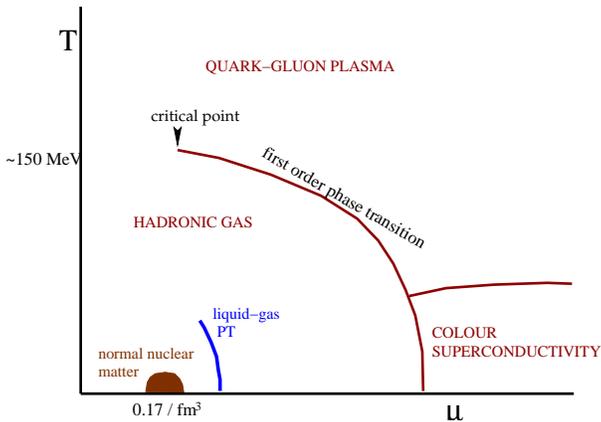}   
\end{center}                      
\vspace{-2mm} \caption{The phase diagram of strongly interacting matter.
\label{phdiag}}
\end{figure}

The phase diagram of strongly interacting matter is depicted in Figure~\ref{phdiag}.
At vanishing and small baryochemical potential hadronic phase changes 
smoothly (though rapidly) into quark-gluon phase and there is no phase transition
(no discontinuity of any derivative of the energy). As the baryochemical potential 
is increased, a first order phase transition appears. The line of the phase transition 
ends in a critical point in which second order phase transition is realised. Its position
is currently unknown and is subject of an intense search. 

As we mentioned, the system expands rather rapidly. If it does reach the quark-gluon phase 
in the region with high baryochemical potential, then it rapidly passes through the boundary 
of the two phases in the phase diagram. It is quite general behaviour that if thermodynamic 
system expands quickly through a phase boundary of \textit{a first order phase transition}
it remains in the high temperature phase and 
supercools. If the expansion is fast enough it reaches the spinodal and fragments 
into pieces of characteristic size. This process is studied in general physics \cite{grady}
but appears also in multifragmentation in nuclear collisions at energies of the order 
100~MeV per nucleon \cite{mfrag,chomaz}. 

\begin{figure} [h,t]                  
\begin{center}                        
\includegraphics[width=70mm]{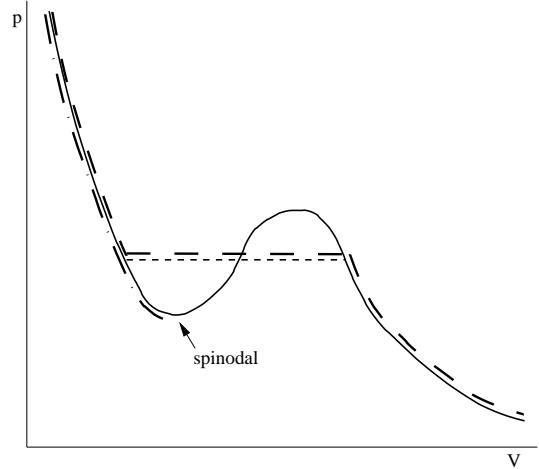}   
\end{center}                      
\vspace{-2mm} \caption{Isotherm of the van der Waals equation of state. Slow 
phase transition follows the Maxwell construction (long-dashed line). 
Fast expansion follows the original van der Waals 
curve until it reaches the local minimum of the isotherm (the spinodal) and fragments
(dash-dotted line).
\label{vdW}}
\end{figure}

On an elementary level spinodal fragmentation can be illustrated with the help of van der Waals 
equation of state (Fig.~\ref{vdW}).
Rapidly expanding system follows the van der Waals curve rather than the Maxwell construction. 
It may follow that curve down to the first minimum; then $dp/dV$ on the isotherm becomes positive, 
which is an instable solution. The system thus goes into spinodal fragmentation. 

This scenario is realised if the expansion rate $V^{-1}\, dV/dt$ is larger than the nucleation rate 
of bubbles of the new phase
$
\Gamma \propto \exp\left ( -{\Delta F_*}/{T} \right ) \,
$,
where $\Delta F_*$ is the difference of free energies: that of the old phase minus free energy of a
bubble of the new phase \cite{kapusta}. Model studies with linear sigma model coupled to quarks
(to model chiral phase transition) indicate that realistic expansion rate indeed is larger than 
the bubble nucleation rate and thus spinodal decomposition is relevant scenario for heavy ion 
collisions \cite{scaven}.

One naively expects that such a decomposition scenario may be disregarded at RHIC and LHC where
the fireball most probably evolves through the rapid crossover transition and not first order 
phase transition (so no spinodal decomposition can be realised). However, it has been noted that 
even in the region of rapid crossover the \textit{bulk}  viscosity $\zeta$ (unlike the 
{\em shear} viscosity) 
suddenly shows a sharp peak as a function of temperature \cite{paech,tuchin}. Recall that 
the corresponding viscous force is proportional to $\zeta \partial_i u^i$ and that in strongly
expansing fireball the divergence of velocity is large. Thus sudden peak in bulk viscosity 
corresponds to following scenario: First the fireball in high-temperature phase begins to 
expand strongly. Then, at crossover suddenly a force appears which basically makes it very stiff
in the sense ``not willing to expand''. This force tends to {\em decelerate} the expansion. 
However, there is inertia of the matter, so it may happen
that the bulk will not be able to respond to the viscous force and fireball will be torn 
apart into pieces. 

Assuming that this kind of breaking happens if the dissipated energy equals kinetic 
energy of the matter,
in \cite{giorgio} it was estimated that  characteristic size of fragments (in one-dimensional 
boost-invariant expansion scenario) is
$
L^2={24\zeta_c\tau_c}/{\varepsilon_c}
$,
where $\tau_c$ is the time when crossover is reached, $\varepsilon_c$ the energy density at that 
point, and $\zeta_c$ is a scaling factor of the bulk viscosity at the same point 
($\zeta(\tau) = \zeta_c \tau_c \delta(\tau - \tau_c)$). 

Such fragmentation would have implications on many observables, mainly correlations and 
fluctuations. A Monte Carlo generator of particles has been developed which simulates particle 
emission from such droplets. 

\begin{figure} [h,t]                  
\begin{center}                        
\includegraphics[width=84mm]{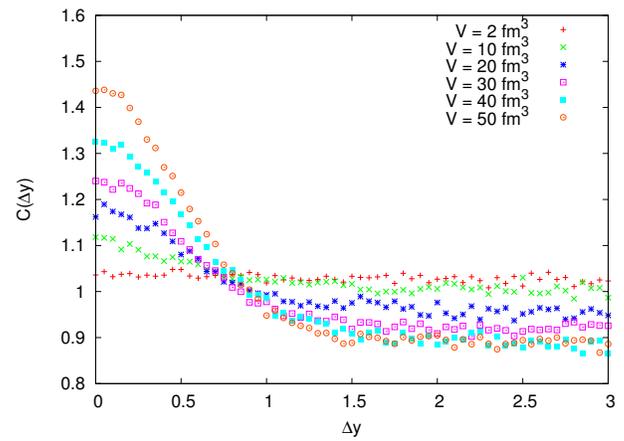}   
\end{center}                      
\vspace{-2mm} \caption{Proton-proton correlation as a function of the rapidity difference. 
No interactions between protons are taken into account, functions are not normalised. Protons 
come from events with 900 hadrons. All hadrons are emitted from droplets. Droplets are 
distributed according to Gaussian with rapidity width of 1.3. Indicated is the volume of fragments.
\label{corf}}
\end{figure}

It has been proposed that emission from droplets will lead to modifications of particle 
correlation functions \cite{pratt,randrup}. Due to their large mass, 
such a modification will be best visible 
for protons \cite{randrup}. We sampled such correlation functions with our Monte Carlo 
generator and observed that the peak of the correlation function clearly appears with 
increasing fraction of particles produced from droplets and increasing droplet size
(Fig.~\ref{corf}). 

Emission of particles from fragments may result in non-statistically varying rapidity 
distributions in individual events. 
We propose a method for recognising non-statistical fluctuations of rapidity distributions
based on Kolmogorov-Smirnov test \cite{melo}. 

A scenario of fireball fragmentation and subsequent emission of particles from fragments
could also help to reconcile hydrodynamic simulations with femtoscopy data. Currently, 
there is a sharp disagreement which does not improve if freeze-out is treated ``correctly''
by using a cascade generator as an afterburner. Major part of the failure is due to the shape 
of freeze-out hypersurface in the simulations. This is modified if our fragmentation scenario
is assumed \cite{giorgio}.

Finally, let us note that there are other effects observed in RHIC data which indicate 
clustering of the hadrons at the emission. PHOBOS collaboration tested such hypotheses
on data on multiplicity fluctuations \cite{phob-mult} and two-particle correlations \cite{phob-tp}. 
It was also conjectured that non-statistical fluctuations of mean $p_t$ at RHIC may be 
due to clustering of particles \cite{bbfh}. 

We shall continue to investigate phenomenological consequences of the fragmentation scenario. 

This research has been supported in parts by VEGA 1/4012/07, MSM 6840770039, and LC 07048.


\end{document}